\documentstyle[12pt]{article}
\evensidemargin\oddsidemargin
\newcommand{\alt}{\mathrel{\raisebox{-.6ex}{$\stackrel{\textstyle<}{\sim}$}}}
\newcommand{\agt}{\mathrel{\raisebox{-.6ex}{$\stackrel{\textstyle>}{\sim}$}}}

\def\overlay#1#2{\ifmmode \setbox 0=\hbox {$#1$}\setbox 1=\hbox to\wd 0{\hss
$#2$\hss }\else \setbox 0=\hbox {#1}\setbox 1=\hbox to\wd 0{\hss #2\hss }\fi
#1\hskip -\wd 0\box 1}

\def\nv#1 {\noalign{\vskip#1pt}}

\def\abstract#1{\begin{center}\Large\bf Abstract\end{center}
{\narrower\small #1\par}}

\def\gev{{\rm\,GeV}}

\def\L{{\cal L}}
\def\O{{\cal O}}
\def\pb{{\rm\,pb}}
\def\fb{{\rm\,pb}}

\begin{document}

\font\fortssbx=cmssbx10 scaled \magstep2
\hbox to \hsize{
\hskip.5in \raise.1in\hbox{\fortssbx University of Wisconsin - Madison}
\hfill$\vcenter{\hbox{\bf MAD/PH/752}
                \hbox{May 1993}}$ }

\vspace{.2in}

\begin{center}
{\LARGE\bf SEARCH FOR PHYSICS\\[.1in]
 BEYOND THE STANDARD MODEL\footnote{Talk presented by V.~Barger at the {\it
International Symposium on 30~Years of Neutral Currents}, Santa Monica,
California (1993)}}\\[.4in]
{\large V.~Barger$^a$  and R.J.N.~Phillips$^b$}\\[.2in]
\it
$^a$Physics Department, University of Wisconsin, Madison, WI 53706, USA\\
$^b$Rutherford Appleton Laboratory, Chilton, Didcot, Oxon OX11 0QX, UK
\end{center}

\renewcommand{\LARGE}{\Large}
\renewcommand{\Huge}{\Large}

\vspace{.25in}

\abstract{
We survey some recent ideas and progress in looking for particle physics
beyond the Standard Model, connected by the theme of Supersymmetry
(SUSY).  We review the success of SUSY-GUT models, the expected
experimental signatures and present limits on SUSY partner particles,
and Higgs phenomenology in the minimal SUSY model.}

\vspace{.5in}

\section{Introduction}

   As we stand at the beginning of 1993, the Standard Model (SM) is in
excellent shape; all its predictions that have been tested have been
verified to high precision.  Important checks remain to be made, however:
the top quark is not yet discovered, the interactions between gauge bosons
are still unmeasured, and the Higgs boson remains a totally unconfirmed
hypothesis.  There may still be some surprises here, especially in the
Higgs sector.

   But even if all these checks give SM results, the apparent
arbitrariness and the theoretical limitations of the SM suggest the
workings of some deeper principles, embodied perhaps in Supersymmetry
(SUSY) or Superstrings.  Such ideas imply new physics, new particles
and new interactions beyond the SM.  The present review covers a small
number of selected topics related to SUSY: unification of couplings in
SUSY-GUT models, experimental signals from SUSY, and Higgs phenomenology
in the minimal SUSY extension of the SM (MSSM).

   SUSY requires each fermion to have a boson partner (and vice versa),
with all the same quantum numbers but with spin differing by 1/2.  Since
no such partners have been found, SUSY is plainly a broken symmetry at
presently explored mass scales but could hold at a higher scale;
we denote the typical scale of the superpartners as $M_{\rm SUSY}$.

   The primary theoretical motivation for SUSY is that it stabilizes divergent
loop contributions to scalar masses, because fermion and boson loops
contribute with opposite signs and largely cancel.  This cures the
naturalness problem in the SM, so long as $M_{\rm SUSY} \alt \O(1\rm\ TeV)$,
where otherwise the Higgs mass would
require fine-tuning of parameters. There are also attractive practical
features:  SUSY-GUT models can be
calculated perturbatively and can be tested experimentally at supercolliders,
where SUSY partners can be produced and studied.  Philosophically, SUSY
is the last possible symmetry of the $S$-matrix~\cite{haag},
and there is a predisposition to believe that anything not forbidden is
compulsory.

   The development of SUSY ideas in recent years is briefly as
follows~\cite{reviews}.

\[\vbox{
\begin{tabular}{ll}
\multicolumn{2}{c}{\bf SUSY TIMELINE}\\ \hline
\\
1966--68: \begin{tabular}[t]{@{}l}SUSY for baryon-meson system;\\
                                  SUSY algebra (non-relativistic)
          \end{tabular}
& Miyazawa\\
\\
1971: \begin{tabular}[t]{@{}l}Two-dimensional supersymmetry:\\
                              dual string models with fermions
      \end{tabular}
& \begin{tabular}[t]{@{}l}Ramond\\
                          Neveu + Schwarz
  \end{tabular}\\
\\
1973: Four-dimensional supersymmetric field theories& Wess + Zumino\\
\\
1974: Absences of many divergences&  Wess + Zumino\\
& Iliopoulos\\
& Ferrara\\
\\
1976: \begin{tabular}[t]{@{}l}Supergravity\\
                             (local supersymmetry)
      \end{tabular}
& \begin{tabular}[t]{@{}l}Friedman + \\
                          van Nieuwenhuizen + Ferrara\\
                          Deser + Zumino
  \end{tabular}\\
\\
1977: Model building& Fayet + \dots\\
\\
1981: Naturalness/hierarchy& Maiani\\
& 't\,Hooft\\
& Witten\\
\\
1981: SUSY SU(5) GUT& Dimopoulos + Raby + Wilczek\\
& Dimopoulos + Georgi\\
& Sakai\\
\\
1984: Strings& Green + Schwarz\\
\\
1987--present: SUSY RGEs + data& Amaldi et al.\\
& + \dots \\
\\ \hline
\end{tabular}
}\]

\newpage\noindent
   The high publication rate for SUSY papers has now reached a plateau,
showing a steady continuation of interest (Fig.~1)

\begin{center}

\parbox{5.5in}{Fig.1.  SUSY papers in SPIRES, 1980--1992, containing
``supersymmetry" or ``supersymmetric" in the title (figure provided by
A.~L.~Stange).}
\end{center}

\medskip

   Phenomenological interest has focussed mainly on the Minimal
Supersymmetric extension of the SM (MSSM), which introduces just one
spartner for each SM particle.  The gauge symmetry is $\rm SU(3)_c\times
SU(2)_L\times U(1)_Y$; the corresponding spin-1 gauge bosons $g,W,Z,\gamma$
have spin-1/2 ``gaugino" partners $\tilde g,\tilde W, \tilde Z, \tilde \gamma$.
The three generations of spin-1/2 quarks $q$ and leptons $\ell$ have spin-0
squark and slepton partners $\tilde q$ and $\tilde\ell$; the
chiral states $f_L$ and $f_R$ of any given fermion $f$ have distinct sfermion
partners $\tilde f_L$ and $\tilde f_R$, respectively.  For anomaly cancellation
the single Higgs doublet must be replaced by two doublets $H_1$
and $H_2$ that have higgsino partners $\tilde H_1$ and $\tilde H_2$.
The MSSM also conserves a multiplicative $R$-parity, defined by
\begin{equation}
    R = (-1)^{2S+L+3B}
\end{equation}
where $S,L,B$ are spin, lepton number and baryon number.  $R$ distinguishes
the normal particles of the SM, which all have $R=+1$, from their spartners
which differ simply by 1/2 unit of $S$ and therefore have $R=-1$.
$R$-conservation comes from  restricting the types of coupling that are
allowed.  It has immediate and important physical implications:
\begin{enumerate}
\item sparticles must be produced in pairs,
\item heavy sparticles decay to lighter sparticles,
\item the lightest sparticle (LSP) is stable.
\end{enumerate}
If this LSP has zero charge and only interacts weakly, as seems likely
since no candidates are yet discovered, it will carry off undetected
energy and momentum in high-energy collisions (providing possible
signatures for sparticle production) and will offer a possible source of
cosmological dark matter.

   As work has proceeded, several significant phenomenological motivations
for SUSY have emerged, in addition to the more general motivations
above.

\begin{enumerate}
\item Grand Unified Theories (GUTs) with purely SM particle content do not
predict a satisfactory convergence of the gauge couplings at some high
GUT scale $M_G$, but convergence can be achieved if SUSY partners are added
(see Section~2)~\cite{amaldi,ellis,langacker}.

\item Starting from equal $b$ and $\tau$ Yukawa couplings at the GUT scale
$M_G$, the physical masses can be correctly predicted when the evolution
equations include SUSY partners, but not with the SM alone (see
Section~2)~\cite{ramond,giveon}.

\item Proton decay is too rapid in a SM GUT but can be acceptable in
SUSY-GUT models where $M_G$ is higher~\cite{hisano}.

\item Assuming $R$-parity conservation, the lightest SUSY partner (LSP) is
stable and provides a plausible candidate for the origin of dark matter
making $\Omega \sim 1$~\cite{drees,kelley,roberts}.

\item SUSY-GUT models lead naturally to the Higgs field developing a vacuum
expectation value, when the top mass is larger than $M_W$~\cite{ross}.
\end{enumerate}

\section{Unification of couplings in SUSY-GUT models}

   The evolution of couplings, as the renormalization mass scale $\mu$
is changed, is governed by the Renormalization Group Equations (RGE).
For the gauge group $\rm SU(3)\times SU(2)\times U(1)$,
with corresponding gauge couplings
$g_3(=g_s), g_2(=g), g_1(=\sqrt{5/3}g')$ , the RGE can be written in
terms of the dimensionless variable $t=\ln(\mu/M_G)$:
\begin{equation}
{dg_i\over dt} = {g_i\over 16\pi^2} \left[ b_ig_i^2 + {1\over16\pi^2}
\left( \sum_{j=1}^3 b_{ij} g_i^2g_j^2 - \sum a_{ij}
g_i^2\lambda_j^2\right)\right]
\end{equation}
The first term on the right is the one-loop approximation; the second
and third terms contain two-loop effects, involving other gauge
couplings $g_j$ and Yukawa couplings $\lambda_j$.  The coefficients $b_i,\
b_{ij}$ and $a_{ij}$ are determined at given scale $\mu$ by the content of
active particles  (those with mass ${}<\mu$).  If there are no thresholds
({\it i.e.} no
changes of particle content) between $\mu$ and $M_G$, then the coefficients
are constants through this range and the one-loop solution is
\begin{equation}
\alpha_i^{-1}(\mu)  =  \alpha_i^{-1}(M_G) - t b_i/(2\pi)   \;,
\end{equation}
where $\alpha_i = g_i^2/(4\pi)$; thus $\alpha_i^{-1}$ evolves linearly with
$\ln\mu$ at one-loop order.    If there are no new physics thresholds
between $\mu = M_Z \simeq m_t$  and $M_G$  ({\it i.e.} nothing but a ``desert"
as in the basic SM) then equations of this kind should evolve the observed
couplings at the electroweak scale~\cite{giatw}
\begin{eqnarray}
\alpha_1(M_Z)^{-1} &=& 58.89 \pm 0.11 \,, \\
\alpha_2(M_Z)^{-1} &=& 29.75 \pm 0.11 \,, \\
\alpha_3(M_Z)^{-1} &=& 0.118 \pm 0.007 \,,
\end{eqnarray}
to converge to a common value at some large scale.  Figure~2 shows that
such a SM extrapolation does NOT converge; this figure actually includes
two-loop effects but the evolution is still approximately linear versus
$\ln\mu$, as at one-loop order.  GUTs do not work, if we assume just SM
particles plus a desert up to $M_G$.

\begin{center}

Fig.~2.  Evolution of gauge couplings in the SM.
\end{center}

   If however we increase the particle content to include the minimum
number of SUSY particles, with a threshold not too far above $M_Z$,  then
GUT-type convergence can happen.  Figure~3 shows two examples with SUSY
threshold $M_{\rm SUSY}= m_t = 150$~GeV or $M_{\rm SUSY}=1$~TeV~\cite{bbo},
the threshold difference being
compensated by a small change in $\alpha_3(M_Z)$.  SUSY-GUTs are plainly
more successful; the evolved couplings are consistent with a common
intersection at $M_G \sim 10^{16}$\,GeV.  In fact a precise single-point
intersection is not strictly necessary, since the exotic GUT gauge,
fermion and scalar particles do not have to be precisely degenerate;
we may therefore have several non-degenerate thresholds near $M_G$, to be
passed through on the way to GUT unification.

\begin{center}
\medskip

\parbox{5in}{Fig.~3.  Evolution of gauge couplings in two SUSY-GUT examples,
with SUSY thresholds at $M_{\rm SUSY}=m_t=150$~GeV or
$M_{\rm SUSY}=1$~TeV~\cite{bbo}.}

\end{center}

   The Yukawa couplings also evolve.  The evolution equations for $\lambda_t$
and $\lambda_b/\lambda_\tau$ are
\begin{equation}
{d\lambda_t\over dt} = {\lambda_t\over16\pi^2} \left[-\sum c_i g_i^2 +
6\lambda_t^2 + \lambda_b^2  + \mbox{2-loop terms}\right] \;, \label{yuklam_t}
\end{equation}
with $c_1=13/15$, $c_2=3$, $c_3=16/3$, and
\begin{equation}
{d(\lambda_b/\lambda_\tau)\over dt} = {(\lambda_b/\lambda_\tau)\over16\pi^2}
\left[-\sum d_i g_i^2+\lambda_t^2+3\lambda_b^2-3\lambda_\tau^2
+ \mbox{2-loop terms} \right]\,, \label{yuklam_b}
\end{equation}
with $d_1=-4/3$, $d_2=0$, $d_3=16/3$. The low-energy values at $\mu=m_t$ are
\begin{eqnarray}
\lambda_b(m_t) &=& {\sqrt2\, m_b(m_b)\over\eta_b v\cos\beta} \;,
\label{lambda_b}\\
\lambda_\tau(m_t) &=& {\sqrt2m_\tau(m_\tau)\over \eta_\tau v\cos\beta} \;, \\
\lambda_t(m_t) &=& {\sqrt2 m_t(m_t)\over v\sin\beta} \;, \label{lambda_t}
\end{eqnarray}
where $\eta_f = m_f(m_f)/m_f(m_t)$ gives the running of the masses below
$\mu=m_t$, obtained from 3-loop QCD and 1-loop QED evolution. The $\eta_q$
values depend on the value of $\alpha_3(M_Z)$, as illustrated in Fig.~4. The
running mass values are $m_\tau(m_\tau)=1.777$~GeV and
$m_b(m_b)=4.25\pm0.15$~GeV.  The denominator factors in
Eqs.~(\ref{lambda_b})--(\ref{lambda_t}) arise from the two Higgs vevs
$v_1=v\cos\beta$ and $v_2=v\sin\beta$; they are related to the SM vev
$v=246$~GeV by $v_1^2+v_2^2=v^2$, while $\tan\beta=v_2/v_1$ measures their
ratio.

\begin{center}

Fig.~4. The scaling factors $\eta_q$ for the running masses
$m_q(\mu)$~\cite{bbo}.
\end{center}

A common boundary condition assumed at the GUT scale is that the $b$-quark and
$\tau$-lepton Yukawa couplings are equal there~\cite{eight,dhr}:
\begin{equation}
\lambda_b(M_G) = \lambda_\tau(M_G) \;. \label{b=tau}
\end{equation}
Figure 5 illustrates the running of $\lambda_t$, $\lambda_b$ and
$\lambda_\tau$, obtained from solutions to the RGEs with the appropriate
low-energy boundary conditions and the GUT-scale condition of~(\ref{b=tau}).
Note that  $\lambda_t(M_G)$ must be large in order to satisfy the  boundary
condition
$m_b(m_b)=4.25\pm0.15$.

\begin{center}

Fig.~5. The running of $\lambda_t$, $\lambda_b$ and $\lambda_\tau$ from low
energies to the GUT scale~\cite{bbo}.
\end{center}

As $\mu\to m_t$, $\lambda_t$ rapidly approaches a fixed point~\cite{pendleton}.
The approximate fixed-point solution for $m_t$ is
\begin{equation}
-\sum c_ig_i^2 + 6\lambda_t^2 + \lambda_b^2 = 0 \,.
\end{equation}
Neglecting $g_1,\, g_2$ and $\lambda_b$, $m_t$ is predicted in terms of
$\alpha_s(m_t)$ and $\beta$~\cite{kelley,dhr,bbhz}:
\begin{eqnarray}
m_t(m_t) &\approx& \frac{4}{3}\sqrt{2\pi\alpha_s(m_t)}\,
\frac{v}{\sqrt2}\sin\beta \nonumber\\
&\approx& \frac{v}{\sqrt2}\sin\beta \nonumber\\
&\approx& (200\gev){\tan\beta\over\sqrt{1+\tan^2\beta}} \,.
\end{eqnarray}
Thus the natural scale of the top-quark mass is large in SUSY-GUT models. Note
that the propagator-pole mass is related to this running mass by
\begin{equation}
m_t({\rm pole}) = m_t(m_t)\left[1+{4\over3\pi}\alpha_s(m_t)+\cdots\right]\,.
\end{equation}

An exact numerical solution for the relation between $m_t$ and $\tan\beta$,
obtained from the 2-loop RGEs for $\lambda_t$ and $\lambda_b/\lambda_\tau$, is
shown in Fig.~6~\cite{bbo} taking $M_{\rm SUSY}=m_t$. At large $\tan\beta$,
$\lambda_b$ becomes large and the above fixed-point solution no longer applies.
In fact, the solutions becomes non-perturbative at large $\tan\beta$ and we
impose the perturbative requirements $\lambda_t(M_G)\leq3.3$,
$\lambda_b(M_G)\leq3.1$, based on the requirement that (2-loop)/(1-loop)$\leq
1/4$. At large $\tan\beta$ there is the possibility of
$\lambda_t=\lambda_b=\lambda_\tau$ unification.
For most $m_t$ values there are two possible solutions for $\tan\beta$; the
lower solution is
\begin{equation}
\sin\beta \simeq m_t({\rm pole})/200\gev \,.
\end{equation}
An upper limit $m_t(\rm pole)\alt200$~GeV is found with the RGE solutions.

\bigskip

\begin{center}

Fig.~6. Contours of constant $m_b(m_b)$ in the
$\left(m_t(m_t),\tan\beta^{\vphantom1}\right)$ plane~\cite{bbo}.
\end{center}

\bigskip

Figure 7 shows the dependence of $\lambda_t(M_G)$ on $\alpha_3(M_Z)$. For
$\lambda_t$ to remain perturbative, an upper limit $\alpha_3(M_Z)\alt0.125$ is
necessary.

\begin{center}

Fig.~7. Dependence of  $\lambda_t$ at the GUT scale on
$\alpha_3(M_Z)$~\cite{bbo}.
\end{center}

\medskip

Specific GUT models also make predictions for CKM matrix elements. For example,
several models~\cite{dhr,hrr} give the GUT-scale relation
\begin{equation}
 |V_{cb}(\rm GUT)|=\sqrt{\lambda_c({\rm GUT})/\lambda_t({\rm GUT})}\,.
\end{equation}
The relevant RGEs are
\begin{eqnarray}
{d|V_{cb}|\over dt} &=&
-{|V_{cb}|\over16\pi^2}\left[\lambda_t^2+\lambda_b^2+\mbox{2-loop}\right]\,,\\
{d(\lambda_c/\lambda_t)\over dt} &=& -{(\lambda_c/\lambda_t)\over16\pi^2}
\left[3\lambda_t^2+\lambda_b^2+\mbox{2-loop}\right]\,,
\end{eqnarray}
in addition to Eqs.~(\ref{yuklam_t}) and (\ref{yuklam_b}).
Starting from boundary conditions on $m_c$ and $|V_{cb}|$ at scale $\mu=m_t$,
the equations can be integrated up to $M_G$ and checked to see if the above
GUT-scale constraint is satisfied. The low-energy boundary conditions are
\begin{equation}
0.032 \le |V_{cb}(m_t)|\le0.054\,, \qquad 1.19\le m_c(m_c)\le1.35\gev \,.
\end{equation}
The resulting solutions at the 2-loop level are shown by the dashed curves in
Fig.~8. The contours of $m_b(m_b)=4.1$ and 4.4~GeV, which satisfy
$\lambda_b(M_G)/\lambda_\tau(M_G)=1$, are also shown. The shaded region in
Fig.~8(a) denotes the solutions that satisfy both sets of GUT-scale
constraints. A lower limit $m_t\ge155$~GeV can be inferred, based on values
$m_c=1.19$ and $\alpha_3(M_Z)=0.110$ in this illustration; with
$\alpha_3(M_Z)=0.118$ instead, $m_t$ can be as low as 120~GeV with
$|V_{cb}|=0.054$. One GUT ``texture'' that leads to the above $|V_{cb}|$ GUT
prediction is given by the following up-quark, down-quark and lepton mass
matrices at $M_G$~\cite{dhr}:
\begin{equation}
{\bf U}= \left(\begin{array}{ccc}
0&C&0\\ C&0&B\\ 0&B&A\end{array}\right) \qquad
{\bf D}= \left(\begin{array}{ccc}
0&Fe^{i\phi}&0\\ Fe^{-i\phi}&E&0\\ 0&0&D\end{array}\right) \qquad
{\bf E}= \left(\begin{array}{ccc}
0&F&0\\ F&E&0\\ 0&0&D\end{array}\right) .
\end{equation}

\begin{center}

\parbox{6.5in}{Fig.~8. Contours of constant $m_b(m_b)$ for
$\lambda_b/\lambda_\tau=1$ at $\mu=M_G$ and contours of constant
$|V_{cb}(m_t)|$, (a)~in the $\left(m_t(m_t),\sin\beta^{\vphantom1}\right)$
plane and (b)~in the $\left(m_t(m_t),\tan\beta^{\vphantom1}\right)$
plane~\cite{bbo,bbhz}.}
\end{center}

\section{Experimental SUSY Signatures}

Experimental evidence for SUSY could come in various forms, for example
\begin{enumerate}
\item discovery of one or more superpartners,
\item discovery of a light neutral Higgs boson with non-SM properties and/or a
charged Higgs boson,
\item discovery of $p\to K\nu$ decay: the present lifetime limit is $10^{32}$
years but Super-Kamiokande will be sensitive up to 10$^{34}$ years,
\item discovery that dark matter is made of heavy ($\alt 100$~GeV) neutral
particles.
\end{enumerate}

GUTs are essential for SUSY phenomenology, since otherwise there would be far
too many free parameters. A minimal set of GUT parameters with soft SUSY
breaking consists of the gauge and Yukawa couplings $g_i$ and $\lambda_i$, the
Higgs mixing mass $\mu$, the common gaugino mass at the GUT scale $m_{1/2}$,
the common scalar mass at the GUT scale $m_0$, and two parameters $A,B$ that
give trilinear and bilinear scalar couplings. At the weak scale, the gauge
couplings are experimentally determined. The Higgs potential depends upon
$m_0,\mu,B$ (at tree level) and $m_{1/2},A,\lambda_t,\lambda_b$ (at one loop).
After minimizing the Higgs potential and putting in the measured $Z$ and
fermion masses, there remain 5 independent parameters that can be taken as
$m_t,\tan\beta,m_0,m_{1/2},A$, though other independent parameter sets are
often used for specific purposes.

The SUSY particle spectrum consists of Higgs bosons $(h,H,A,H^\pm)$, gluinos
$(\tilde g)$, squarks $(\tilde q)$, sleptons $(\tilde\ell^\pm)$, charginos
($\tilde W_i^\pm, i=1,2$; mixtures of winos and charged higgsinos), neutralinos
($\tilde Z_j, j=1,2,3,4$; mixtures of zinos, photinos and neutral higgsinos).
An alternate notation is $\tilde\chi_i^+$ for $\tilde W_i^+$ and
$\tilde\chi_j^0$ for $\tilde Z_j$. The evolution of the SUSY mass spectrum from
the GUT scale~\cite{ross,rosrob} is illustrated in Fig.~9. The running masses
are plotted
versus
$\mu$ and the physical value occurs where the running mass $m=m(\mu)$
intersects the curve $m=\mu$. In the case of the Higgs scalar $H_2$, the
mass-square becomes negative at low $\mu$ due to coupling to top; in this
region we have actually plotted $-|m(\mu)|$. Negative mass-square parameter is
essential for spontaneous symmetry-breaking, so this feature of SUSY-GUTs is
desirable; here it is achieved by radiative effects. The running masses for the
gauginos $\tilde g,\tilde W,\tilde B$ are given by
\begin{equation}
M_i(\mu)=m_{1/2}\,{\alpha_i(\mu)\over\alpha_i(M_G)} \,,
\end{equation}
where $i$ labels the corresponding gauge symmetry; this applies before we add
mixing with higgsinos to obtain the chargino and neutralino mass eigenstates.
In the example of Fig.~9 the squarks are heavier than the gluinos, but the
opposite ordering $m_{\tilde q}< m_{\tilde g}$ is possible in other scenarios.
Sleptons, neutralinos and charginos are lighter than both squarks and gluinos
in general.
Note that the usual soft SUSY-breaking mechanisms preserve the
gauge coupling relations (unification) at $M_G$.

\begin{center}

Fig.~9. Representative RGE results for spartner masses~\cite{ross}.
\end{center}

   In order that SUSY cancellations shall take effect at low mass
scales as required, the SUSY mass parameters are expected to be
bounded by
\begin{equation}
      m_{\tilde g},\, m_{\tilde q},\, |\mu|,\, m_A \alt 1\mbox{--2 TeV}\, .
\label{m bound}
\end{equation}
The other parameter $\tan\beta$ is effectively bounded by
\begin{equation}
          1  \alt    \tan\beta  \alt    65    \;,
\end{equation}
where the lower bound arises from consistency in GUT models and the
   upper bound is the perturbative limit. Proton decay gives the constraint
$\tan\beta<85$~\cite{hisano}.

      At LEP\,I, sufficiently light SUSY particles would be produced
through their gauge couplings to the $Z$.  Direct searches for
SUSY particles at LEP give mass lower bounds
\begin{equation}
      m_{\tilde q},\, m_{\tilde \ell},\,m_{\tilde\nu},\,
 m_{\tilde W_1}\agt\mbox{40--45 GeV}\;.
\end{equation}
The limitation of LEP is its relatively low CM energy.

      Hadron colliders can explore much higher energy ranges. Figure~10
shows lowest-order gluon-gluon, gluon-quark and quark-antiquark
subprocesses for SUSY particle hadroproduction.  Figure~11 shows
squark and gluino predictions for the Tevatron $p$-$\bar p$
collider~\cite{bt,btw},
assuming degenerate masses $m_{\tilde q} = m_{\tilde g}$ (summing
$L$ and $R$ squarks plus antisquarks of all flavors).  The right-hand
vertical axis shows the number of events for the luminosity 25~pb$^{-1}$
expected in 1993; we see that about 100 events would be expected
for each of the channels $\tilde g\tilde q$ and $\tilde q\tilde q$
at mass 200~GeV, so the Tevatron clearly reaches well beyond the
LEP range.

\begin{center}

Fig.~10. Typical SUSY production subprocesses in hadron collisions.

\bigskip

\smallskip

\parbox{4.5in}{Fig.~11. Tevatron cross sections for $\tilde g\tilde g$, $\tilde
g\tilde q$ and $\tilde q\tilde q$ production, versus squark/gluino mass.}
\end{center}

\medskip

      The most distinctive signature of SUSY production is the missing
   energy and momentum carried off by the undetected LSP, usually
   assumed to be the lightest neutralino $\tilde Z_1$, which occurs in
   all SUSY decay chains with $R$-parity conservation.  At hadron
   colliders
   it is only possible to do book-keeping on the missing transverse
   momentum denoted $\overlay/p_T$.   The missing momenta of both LSPs are
added
   vectorially in $\overlay/p_T$.  The LSP momenta and hence the magnitude
   of  $\overlay/p_T$ depend on the decay chains.

      If squarks and gluinos are rather light ($m_{\tilde g},m_{\tilde q}\alt
50$~GeV), their dominant decay
   mechanisms are direct strong decays or decays to the LSP:
\begin{eqnarray}
\left.\begin{array}{l}
      \tilde q \to q \tilde g\\
\tilde g \to q \bar q \tilde Z_1
      \end{array} \right\}
&& {\rm if}\ m_{\tilde g} < m_{\tilde q}  \label{light a}\\
\left.\begin{array}{l}
           \tilde g\to q \tilde q \phantom{Z_1} \\
\tilde q\to  q \tilde Z_1  \phantom{q}
\end{array}\right\}
&&    {\rm if}\ m_{\tilde q} < m_{\tilde g} \label{light b}
\end{eqnarray}
   In such cases the LSPs carry a substantial fraction of the available
   energy and $\overlay/p_T$ is correspondingly large.  Assuming such decays
and small LSP mass, the present 90\%~CL experimental bounds from UA1 and UA2
   at the CERN $p$-$\bar p$ collider ($\sqrt s = 640$~GeV) and from CDF at the
   Tevatron ($\sqrt s = 1.8$~TeV) are~\cite{uacdf}
\[
\vbox{\tabskip2em\halign{#\hfil&&$#$\hfil\cr
                    &  \hfil  m_{\tilde g}  &  \hfil  m_{\tilde q}\cr
        UA1 (1987)      &     >  53\rm\ GeV  &      >  45\rm\ GeV \cr
        UA2 (1990)      &        >  79       &     >  74 \cr
        CDF (1992)      &        > 141       &     > 126 \cr}}
\]
   The limits become more stringent if the squark and gluino masses
   are assumed to be comparable.

   For heavier gluinos and squarks, many new decay channels are open,
such as decays into the heavier gauginos:
\begin{eqnarray}
  \tilde g  &\to& q \bar q \tilde Z_i\ (i=1,2,3,4),\ q \bar q' \tilde W_j\
                 (j=1,2),\ g \tilde Z_1 \;, \label{heavy a}\\
  \tilde q_L &\to& q \tilde Z_i\ (i=1,2,3,4),\ q' \tilde Wj\ (j=1,2)\;,
\label{heavy b}\\
  \tilde q_R &\to& q \tilde Z_i\ (i=1,2,3,4)\;. \label{heavy c}
\end{eqnarray}
Some decays go via loops (e.g.\ $\tilde g\to g \tilde Z_1)$; we have not
attempted an exhaustive listing here.
Figure~12 shows how gluino-to-heavy-gaugino branching fractions increase
with $m_{\tilde g}$ in a particular example
(with $m_{\tilde g} < m_{\tilde q}$)~\cite{bbkt}.

   The heavier gauginos then decay too:
\begin{eqnarray}
  \tilde W_j &\to& Z \tilde W_k,\, W \tilde Z_i,\, H_i^0 \tilde W_k,\,
                 H^\pm \tilde Z_i,\,f\tilde f   \;, \\
  \tilde Z_i &\to& Z \tilde Z_k,\, W \tilde W_j,\, H_i^0 \tilde Z_k,\,
                 H^\pm \tilde W_k,\, f\tilde f' \;.   \label{Ztwid decay}
\end{eqnarray}
Here it is understood that final $W$ or $Z$ may be off-shell and
materialize as fermion-antifermion pairs; also $Z$ may be replaced
by $\gamma$.  In practice, chargino decays are usually dominated by
$W$-exchange transitions (Fig.~13a); neutralino decays are often dominated
by sfermion exchanges (Fig.~13b) because the $\tilde Z_2 \tilde Z_1 Z$
coupling is small.  To combine the complicated production and cascade
possibilities systematically, all these channels have been incorporated in
the ISAJET~7.0 Monte Carlo package called ISASUSY~\cite{bppt}.

\begin{center}

Fig.~12. Example of gluino decay branchings versus mass~\cite{bbkt}.
\end{center}

\begin{center}

\bigskip
\parbox{4.5in}{Fig.~13. Examples of (a) chargino decay by $W$-exchange,
(b)~neutralino decay by sfermion exchange.}
\end{center}

   These multibranch cascade decays lead to higher-multiplicity final
states in which the LSPs $\tilde Z_1$  carry a much smaller share of the
available energy, so $\overlay/p_T$ is smaller and less distinctive (Fig.~14),
making detection via $\overlay/p_T$ more difficult.  (Leptonic $W$ or $Z$
decays,
$\tau$ decays, plus semileptonic $b$ and $c$ decays, all give background events
with genuine $\overlay/p_T$; measurement uncertainties also contribute fake
$\overlay/p_T$ backgrounds.)  Experimental bounds therefore become weaker when
we take account of cascade decays.  Figure~15 shows typical CDF 90\% CL
limits in the $(m_{\tilde g}, m_{\tilde q})$ plane; the dashed curves are
limits assuming only direct decays (\ref{light a})--(\ref{light b}), while
solid curves are less restrictive limits including cascade decays~(\ref{heavy
a})--(\ref{Ztwid decay}).

The cascade decays also present new opportunities for SUSY detection.
   Same-sign dileptons (SSD) are a very interesting signal~\cite{bkp}, which
arises naturally from $\tilde g \tilde g$  and  $\tilde g \tilde q$  decays
because of the Majorana character of gluinos, with very
little background. Figure~16 gives an example of this signal.
Eqs.~(\ref{heavy a})--(\ref{Ztwid decay})  show how a heavy gluino or squark
can decay to a chargino
$\tilde W_j$ and hence, via a real or virtual $W$, to an isolated charged
lepton.  For such squark pair decays the two charginos --- and hence the two
leptons --- are constrained to have opposite signs, but if a gluino is
present it can decay equally into either sign of chargino and lepton
because it is a Majorana fermion.  Hence  $\tilde g \tilde g$  or
$\tilde g \tilde q$  systems can decay to isolated SSD plus jets plus
$\overlay/p_T$.  The cascade decays of $\tilde q\tilde q$ via the heavier
neutralinos $\tilde Z_i$ offer similar possibilites for SSD, since the $\tilde
Z_i$ are also Majorana fermions. Cross sections for the Tevatron are
illustrated in Fig.~17.

\begin{center}

\parbox{5.5in}{Fig.~14. Typical $\overlay/p_T$ distributions from direct and
cascade decays of gluino pairs at the Tevatron~\cite{btw}.}

\bigskip

\parbox{5.5in}{Fig.~15. 1992 CDF limits in the $(m_{\tilde g}, m_{\tilde q})$
plane, with or without cascade decays, for a typical choice of
parameters~\cite{btw,uacdf}. }
\end{center}

   Genuinely isolated SSD backgrounds come from the production of $WZ$ or
$Wt\bar t$ or $W^+W^+$ ({\it e.g.}\ $uu \to ddW^+W^+$ by gluon exchange), with
cross sections of order  $\alpha_2^2$   or   $\alpha_2\alpha_3^2$    or
$\alpha_2^2\alpha_3^2$  compared to $\alpha_3^2$ for gluino pair production,
so we expect to control them with suitable cuts.  Very large   $b \bar b$
production gives SSD via semileptonic $b$-decays plus $B$-$\bar B$ mixing, and
also via combined $b\to c\to s \,\ell^+ \nu$   and  $\bar b\to \bar c\,\ell^+
\nu$ decays, but both leptons are produced in jets and can be suppressed by
stringent isolation criteria.  Also $t\bar t$ gives SSD via $t\to b\, \ell^+
\nu$ and  $\bar t\to \bar b\to \bar c\, \ell^+ \nu$, but the latter lepton is
non-isolated.
So SSD provide a promising SUSY signature.

\begin{center}

Fig.~16. Example of same-sign dilepton appearance in gluino-pair decay.

\bigskip


Fig.~17. Same-sign dilepton signals at the Tevatron~\cite{baerev}.
\end{center}

   Gluino production rates at SSC/LHC are much higher than at the Tevatron. At
$\sqrt s=40$~TeV, the cross section is
\begin{equation}
\sigma(\tilde g \tilde g) = 10^4,\,700,\, 6\;\mbox{fb\quad for }
 m_{\tilde g} = 0.3,\,1,\,2\rm\; TeV\;.
\end{equation}
Many different SUSY signals have been evaluated, including
$\overlay/p_T + n\,$jets, $\overlay/p_T +{}$SSD, $\overlay/p_T + n\,$isolated
leptons, $\overlay/p_T + {}$one isolated lepton${}+ Z$, $\overlay/p_T + Z$,
$\overlay/p_T + Z + Z$.
SSC cross sections for some of these signals from  $\tilde g \tilde g$
production are shown versus $m_{\tilde g}$ in Fig.~18 (for two scenarios,
after various cuts);  the labels 3,4,5 refer to numbers of isolated
leptons~\cite{btw}.

   Heavy gluinos can also decay copiously to $t$-quarks~\cite{btw,bps}:
\begin{equation}
\tilde g \to t \bar t\tilde Z_i , t \bar b \tilde W^-, b \bar t
\tilde W^+ \;.
\end{equation}
$t\to bW$  decay then leads to multiple $W$ production.  For example, for a
gluino of mass 1.5~TeV,  the  $\tilde g\to W,\, WW,\, WWZ,\, WWWW$ branching
fractions are typically of order 30\%, 30\%, 6\%, 6\%, respectively. Figure~19
illustrates SSC cross sections for multi-$W$ production via gluino pair
decays (assuming $m_{\tilde g} < m_{\tilde q}$).  We see that
for $m_{\tilde g}\sim 1$~TeV the SUSY rate for $4W$ production can greatly
exceed the dominant SM $4t\to 4W$ mode, offering yet another signal for
SUSY~\cite{bps}.

\begin{center}

Fig.~18. SSC cross sections for various SUSY signals, after cuts~\cite{btw}.
\end{center}

\begin{center}

\parbox{4.5in}{Fig.~19. Typical SSC rates for gluino pair production and decay
to multi-$W$ final states~\cite{bps}.}
\end{center}

To summarize this section:
\begin{enumerate}
\item Experimental SUSY particle searches have hitherto been based
largely on $\overlay/p_T$ signals.  But for $m_{\tilde g},\,m_{\tilde q}
> 50$~GeV cascade decays become important; these cascades both weaken the
simple $\overlay/p_T$
 signals and provide new signals such as same-sign dileptons, which will be
pursued at the Tevatron.
\item For even heavier squarks and gluinos, the cascade decays dominate
completely and provide further exotic (multi-$W,Z$ and multi-lepton)
signatures, which will be pursued at the SSC and LHC.
\item Gluinos and squarks in the expected mass range of Eq.~(\ref{m bound})
will not escape detection.
\end{enumerate}

\section{SUSY Higgs Phenomenology}

In minimal SUSY, two Higgs doublets $H_1$ and $H_2$ are needed to cancel
anomalies and at the same time give masses to both up- and down-type quarks.
Their vevs are $v_1=v\cos\beta$ and $v_2=v\sin\beta$ as mentioned previously.
There are therefore 5 physical scalar states: $h$ and $H$ (neutral CP-even with
$m_h<m_H$), $A$ (neutral CP-odd) and $H^\pm$. At tree level the scalar masses
and couplings and an $h$-$H$ mixing angle $\alpha$ are all determined by two
parameters, conveniently chosen to be $m_A$ and $\tan\beta$. At tree level the
masses obey $m_h\le M_Z,m_A; m_H\ge M_Z,m_A; m_{H^\pm}\ge M_W,m_A$.

Radiative corrections are important, however~\cite{rad}. The most important new
parameters entering here are the $t$ and $\tilde t$ masses; we neglect for
simplicity some other parameters related to squark mixing. One-loop corrections
give $h$ and $H$ mass shifts of order $\delta m^2\sim G_F\,m_t^4\ln(m_{\tilde
t}/m_t)$, arising from incomplete cancellation of $t$ and $\tilde t$ loops. The
$h$ and $H$ mass bounds get shifted up and for the typical case $m_t=150$~GeV,
$m_{\tilde t}=1$~TeV we get
\begin{equation}
m_h < 116\gev < m_H \,.
\end{equation}
There are also corrections to cubic $hAA,\,HAA,\,Hhh$ couplings, to $h$-$H$
mixing, and smaller corrections to the $H^\pm$ mass. Figure~20 illustrates the
dependence of $m_h$ and $m_H$ on $m_A$ and $\tan\beta$, for two different
values of $m_t$ (with $m_{\tilde t}=1$~TeV still). We shall assume $\tan\beta$
obeys the GUT constraints $1\le\tan\beta\le65$ of Eq.~(24).

At LEP\,I, the ALEPH, DELPHI, L3 and OPAL collaborations~\cite{LEP} have all
searched for the processes
\begin{equation}
e^+e^- \to Z \to Z^*h,Ah \,,
\end{equation}
with $Z^*\to\ell\ell,\nu\nu,jj$ and $h,A\to\tau\tau,jj$ decay modes. The $ZZh$
and $ZAh$ production vertices have complementary coupling-strength factors
$\sin(\beta-\alpha)$ and $\cos(\beta-\alpha)$, respectively, helping to give
good coverage. The absence of signals excludes regions of the $(m_A,\tan\beta)$
plane; Fig.~21 shows typical boundaries for various $m_t$ values, deduced from
ALEPH results~\cite{bcps,LEP}. These results imply lower bounds
\begin{equation}
m_h,m_A\agt20\mbox{--45~GeV (depending on}\tan\beta) \,.
\end{equation}
Null searches for $e^+e^-\to H^+H^-$ also exclude a region with
$\tan\beta<1$~\cite{diaz}.

\begin{center}

\medskip
\parbox{5in}{Fig.~20. Contours of $h$ and $H$ masses in the $(m_A,\tan\beta)$
plane for (a)~$m_t=150$~GeV, (b)~$m_t=200$~GeV, with $m_{\tilde t}=1$~TeV.}
\end{center}

\bigskip

\begin{center}

\medskip
\parbox{5in}{Fig.~21. Limits from ALEPH searches for (a)~$Z\to Z^*h$ and
(b)~$Z\to Ah$ at LEP\,I, for various $m_t$ values with $m_{\tilde
t}=1$~TeV~\cite{bcps,LEP}.}
\end{center}

LEP\,II will have higher energy and greater reach. Figure~22 shows approximate
discovery limits in the  $(m_A,\tan\beta)$ plane for various $m_t$ values,
based on projected searches for $e^+e^-\to ZH\to \ell\ell jj,\nu\nu jj,jjjj$
and for $e^+e^-\to (Zh,Ah)\to \tau\tau jj$, assuming energy $\sqrt s=200$~GeV
and luminosity ${\cal L}=500\rm\,pb^{-1}$. $H^\pm$ searches will not extend
this reach.

\begin{center}

\medskip
\parbox{5in}{Fig.~22. Projected limits for various LEP\,II searches, assuming
$\sqrt s=200$~GeV and ${\cal L}=500\rm\,pb^{-1}$~\cite{bcps}.}
\end{center}

Searches for neutral scalars at SSC and LHC will primarily be analogous to SM
Higgs searches:
\begin{enumerate}

\item untagged $\gamma\gamma$ signals from $pp\to(h,H,A)\to\gamma\gamma$ via
top quark loops (Fig.~23);

\item tagged $\gamma\gamma$ signals from $pp\to(h,H,A)\to\gamma\gamma$ plus
associated $t\bar t$ or $W$, permitting lepton tagging via $t\to W\to \ell\nu$
or $W\to\ell\nu$ decays (Fig.~24);

\item ``gold-plated'' four-lepton signals from $pp\to(h,H)\to ZZ$ or
$Z^*Z\to\ell^+\ell^+\ell^-\ell^-$ (Fig.~25).
\end{enumerate}

\noindent
Though qualitatively similar to SM signals, these will generally be smaller due
to the different coupling constants that depend on $\beta$ and $\alpha$.

\bigskip

\begin{center}

Fig.~23. Typical diagram for untagged Higgs${}\to\gamma\gamma$ signals.

\medskip


Fig.~24. Typical diagrams for lepton-tagged Higgs${}\to\gamma\gamma$ signals.

\medskip


Fig.~25. Typical diagrams for ``gold-plated'' four-lepton Higgs signals.
\end{center}

\medskip
For charged Higgs scalars, the only copious hadroproduction source appears to
be top production with $t\to bH^+$ decay (that requires $m_{H^\pm}<m_t-m_b$).
The subsequent $H^+\to c\bar s,\nu\bar\tau$ decays are most readily detected in
the $\tau\nu$ channel (favored for $\tan\beta>1$), with $\tau\to\pi\nu$ decay
(Fig.~26).

\begin{center}

Fig.~26. Typical diagram for $\tau$ signals from top decay via charged-Higgs
modes.
\end{center}
\bigskip

SM $t$-decays give equal probabilities for $e,\nu,\tau$ leptons via $t\to bW\to
b(e,\mu,\tau)\nu$, but the non-standard $t\to bH^+\to b\tau\nu$ leads to
characteristic excess of $\tau$. The strategy is to tag one top quark via
standard $t\to bW\to b\ell\nu$ decay and to study the $\tau/\ell$ ratio in the
associated top quark decay ($\ell=e$ or $\mu$).

Several groups have studied the detectability of these various signals at
SSC/LHC, and they all reach broadly similar
conclusions~\cite{bcps,baer,gunion,kunszt}. Figures~27 and 28 show typical
limits of detectability for untagged and lepton-tagged $\gamma\gamma$ signals
at SSC, assuming luminosities ${\cal L}=20\,\rm fb^{-1}$ (two years of running)
and $m_t=150$~GeV. Figure~29 shows a similar limit for the $H\to4\ell$ search
(no $h\to4\ell$ signal is detectable).  Figure~30 shows typical limits for
detecting the $t\to H^+\to{}$excess $\tau$ signal; here the value of $m_t$ is
critical, since only the range $m_{H^+}<m_t-m_b$ can contribute at all. Putting
all these discovery regions together with the LEP\,I and LEP\,II regions, we
see that very considerable coverage of the $(m_A,\tan\beta)$ plane can be
expected --- but there still remains a small inaccessible region; see Fig.~31.
For $m_t=120$~GeV the inaccessible region is larger, for $m_t=200$~GeV it is
smaller.

\begin{center}

\medskip
\parbox{5.25in}
{Fig.~27. Limits of detectability for $H,h,A$ untagged $\gamma\gamma$ signals
at the SSC, for $\L=20\fb^{-1}$~\cite{bcps}.}
\bigskip


\parbox{5in}%
{Fig.~28. Detectability limits for $H,h$ lepton-tagged signals at the SSC, for
$\L=20\pb^{-1}$~\cite{bcps}.}

\medskip


{Fig.~29. Detectability limit for $H\to4\ell$ signals at the SSC for
$\L=20\fb^{-1}$~\cite{bcps}.}

\medskip


Fig.~30. Detectability limits for $t\to bH^+\to b\tau^+\nu$ signals at the SSC
(from Ref.~\cite{bcps}).


\medskip
\parbox{5.5in}{Fig.~31. Combined LEP and SSC discovery regions for
$m_t=150$~GeV from Ref.~\cite{bcps}; similar results are obtained by other
groups~\cite{baer,gunion,kunszt}.}

\end{center}

Figure 32 shows how many of the MSSM scalars $h,H,A,H^\pm$ would be detectable,
in various regions of the $(m_A,\tan\beta)$ plane. In many regions two or more
different scalars could be discovered, but for large $m_A$ only $h$ would be
discoverable; in the latter region, the $h$ couplings all reduce to SM
couplings, the other scalars become very heavy and approximately degenerate,
and the MSSM essentially behaves like the SM.

\begin{center}

Fig.~32.  How many MSSM Higgs bosons may be discovered (from Ref.~\cite{bcps}).
\end{center}

An indirect constraint on the MSSM Higgs sector is provided by the CLEO bound
on $b\to s\gamma$ decays~\cite{cleo},
\begin{equation}
B(b\to s\gamma)<8.4\times10^{-4}\quad (95\%\ \rm C.L.)  \,.
\end{equation}
In the SM this decay proceeds via a $W$ loop process, but in models with more
than one Higgs doublet there are charged Higgs contributions too (Fig.~33). In
the MSSM both the $W$ and $H$ amplitudes have the same sign and the branching
fraction is directly related to $m_{H^+}$ and $\tan\beta$ (Fig.~34); hence the
CLEO result implies a lower bound on $m_{H^+}$ for given $\tan\beta$ (Fig.~35).
It was recently pointed out~\cite{hewett,bbp} that this CLEO-based constraint
falls in a very interesting and sensitive region when translated to the
$(m_A,\tan\beta)$ plane; see Fig.~36. Taken at face value, it appears to
exclude a large part of the LEP\,II discovery region and furthermore to exclude
much of the otherwise inaccessible region too; with future improvements in the
CLEO bound, perhaps the whole of the inaccessible region could be excluded.

\begin{center}

\medskip

Fig.~33. $W$ and charged-Higgs loop diagrams contributing to $b\to s\gamma$
decays.


\medskip

\parbox{4.5in}{Fig.~34. Dependence of $B(b\to s\gamma)$ on $m_{H^\pm}$ and
$\tan\beta$ in the MSSM (from Ref.~\cite{bbp}), neglecting other SUSY loops.}

\medskip


\parbox{5in}{Fig.~35. Lower bound on $m_{H^+}$ for given $\tan\beta$, from
$b\to s\gamma$ constraint~\cite{bbp}. The region excluded by the CLEO
experimental bound is to the left of the $b\to s\gamma$ curve.}


\medskip

\parbox{4.5in}{Fig.~36. Comparison of $b\to s\gamma$ bound with other MSSM
Higgs constraints in the $(m_A,\tan\beta)$ plane, for
$m_t=150$~GeV~\cite{bbp}.}

\end{center}

It is premature however to reach any firm conclusions from the results above.
The calculations of Ref.~\cite{bbp} are based on the approximation of
Ref.~\cite{grinstein}, but later work indicates possible further small
corrections~\cite{misiak}. More importantly, other SUSY loop diagrams
(especially chargino loops) can give additional contributions of either sign,
leading to potentially significant changes in the
amplitude~\cite{bertolini,barbieri}. However, as theoretical constraints on
SUSY particles become more extensive, and as the $B(b\to s\gamma)$ bound itself
becomes stronger, we may expect this approach to give a valuable constraint in
the MSSM Higgs phenomenology. [Postscript: at the Washington APS meeting April
1993, CLEO reported an improvement
in the bound of Eq.~(38) to $5.4\times10^{-4}$].

Finally we may ask what a future $e^+e^-$ collider could do. We have seen that
part of the MSSM parameter space is inaccessible to $e^+e^-$ collisions at
$\sqrt s=200$~GeV, $\L=500\pb^{-1}$, for $m_t=150$~GeV and $m_{\tilde
t}=1$~TeV. But a possible future linear collider with higher energy and
luminosity could in principle cover the full parameter space. In is interesting
to know what are the minimum $s$ and $\L$ requirements for complete coverage,
for given $m_t$. This question was answered in Ref.~\cite{nolose}, based on the
conservative assumption that only the channels
$e^+e^-\to(Zh,Ah,ZH,AH)\to\tau\tau jj$ would be searched, with no special
tagging. The results are shown in Fig.~37. We have estimated that including all
$Z\to\ell\ell,\nu\nu,jj$ and $h,H,A\to bb,\tau\tau$ decay channels plus
efficient $b$-tagging could increase the net signal $S$ by a factor~6 and the
net background $B$ by a factor~4, approximately; this would increase the
statistical significance $S/\sqrt B$ by a factor~3 and hence reduce the
luminosity requirement by a factor~9 or so. In this optimistic scenario, the
luminosity axis in Fig.~37 would be rescaled downward by an order of magnitude.

\begin{center}

\medskip

\parbox{5in}{Fig.~37. Minimal requirements for a ``no-lose'' MSSM Higgs search
at a future $e^+e^-$ collider. Curves of minimal $(\sqrt s,\L)$ pairings are
shown for $m_t=120$, 150, 200~GeV; the no-lose region for $m_t=150$~GeV is
unshaded~\cite{nolose}.}
\end{center}

\bigskip

To summarize this Section:

\begin{enumerate}

\item The MSSM Higgs spectrum is richer but in some ways more elusive than the
SM case.

\item At least one light scalar is expected.

\item As $m_A\to\infty$ this light scalar behaves like the SM scalar and the
others become heavy.

\item LEP\,I, LEP\,II and SSC/LHC will give extensive but not quite complete
coverage of the MSSM parameter space.

\item For some parameter regions, several different scalars are detectable, but
usually one or more remain undetectable.

\item The $b\to s\gamma$ bound has the potential to exclude large areas of
parameter space (possibly including the inaccessible region) but is presently
subject to some uncertainty.

\item A higher-energy $e^+e^-$ collider could cover the whole MSSM parameter
space, discovering at least the lightest scalar $h$.

\end{enumerate}

\section*{Acknowledgements}

We thank H.~Baer, M.~Berger, and P.~Ohmann for valuable contributions to the
contents of this review. This work was supported in part by the University of
Wisconsin Research Committee with funds granted by
the Wisconsin Alumni Research Foundation, in part by the U.S.~Department of
Energy under contract no.~DE-AC02-76ER00881, and in part by the Texas National
Laboratory Research Commission under grant no.~RGFY9273.

\end{document}